# Actuation and mapping of SAW-induced high-frequency wavefields on suspended graphene membranes

*Hande N. Açıkgöz[1], Dong Hoon Shin[2], Inge C. van der Knijff[1], Allard J. Katan[3], Xiliang Yang[1], Peter G. Steeneken[1], Gerard J. Verbiest[1,\*], Sabina Caneva[1,\*]*

[1]Department of Precision and Microsystems Engineering, Delft University of Technology, Mekelweg 2, 2628 CD Delft, The Netherlands

[2]Department of Electronics and Information Engineering, Korea University, Sejong 30019, Republic of Korea

[3]Kavli Institute of Nanoscience Delft, Lorentzweg 1, 2628 CJ Delft, The Netherlands

*E-mail: s.caneva@tudelft.nl

g.j.verbiest@tudelft.nl

KEYWORDS Surface acoustic waves, graphene, atomic force microscopy, acoustics

ABSTRACT

High frequency acoustic devices based on two-dimensional (2D) materials are unique platforms to design and manipulate the spatiotemporal response of acoustic waves for next-generation sensing and contactless actuation applications. Conventional methods for actuating suspended membranes, however, cannot be applied to all 2D materials, or are limited in frequency. There is, therefore, a need for a universal high-frequency, on-chip actuation technique that can be applied to all types of membranes. Here, we demonstrate that surface acoustic waves (SAWs) can be used to efficiently actuate suspended 2D materials by exciting suspended graphene membranes with high-frequency (375 MHz) Rayleigh surface waves and mapping the resulting vibration field with atomic force acoustic microscopy (AFAM). Acoustic waves travelling from supported to suspended graphene experience a reduction in acoustic wavelength from 10 μm to ~2 μm due to the decrease in effective bending rigidity, leading to a decrease in wave velocity on suspended graphene. By varying the excitation frequency, we observed a change in phase velocity from ~160 m/s to ~700 m/s. This behavior is consistent with the nonlinear dispersion of acoustic waves, as predicted by plate theory, in suspended graphene membranes. The geometry and bending rigidity of the membrane thus play key roles in modulating the acoustic wave pattern and wavelength. This combined SAW actuation and AFAM visualization scheme can give new insights into the fundamentals of acoustic transport at the nanoscale limit and provides a route towards the manipulation of localized wavefields for on-chip patterning and transport over 2D materials surfaces.

The dynamics of two-dimensional (2D) materials hold significant importance in the fields of electronics, optoelectronics, and micro/nano-electromechanical systems (MEMS/NEMS).[1,2] These materials exhibit unique mechanical, electrical, and thermal properties not observed in their bulk counterparts, and due to their atomic-level thickness, they exhibit enhanced sensitivity to external stimuli.[3,4] The dynamic behavior in 2D materials is typically investigated by actuating a suspended membrane with an external stimulus and analyzing the resultant vibrations. Commonly employed actuation methods include electrostatic, opto-mechanical (thermal), and base-excitation techniques, each offering distinct advantages and limitations.[5] Electrostatic actuation can achieve high frequencies up to the gigahertz (GHz) range[6] and is easily integrated into MEMS/NEMS devices. However, it suffers from non-linear dependence on the applied alternating current (AC) voltage and is restricted to conductive membrane materials. Photothermal or optical actuation can reach extremely high frequencies, into the terahertz (THz) range,[7,8] but involves complex setups and poses challenges for integration with other modules, such as microscopy. Mechanical base-excitation can be applied via piezoelectric transducers which provide high precision and on-chip capabilities[9,10] with frequencies up to 100 MHz [11] while being purely mechanical with no material limitations on suspended layers. Nonetheless, operating these transducers at higher frequencies is difficult and coupling of acoustic power from the bulk of a chip into the suspended membrane is inefficient and hard to control.

Given these constraints, an ideal actuation method for suspended 2D material membranes should achieve high frequencies, enable efficient and controlled on-chip actuation, and not be limited by the membrane's properties (*e.g.* electrical conductivity). Here, surface acoustic waves (SAWs) emerge as a promising candidate. They can reach frequencies in the GHz range,[12,13] enabling on-chip actuation, and relying on a purely mechanical interaction, they are applicable to all membrane types. Confinement of the elastic wave on the surface allows effective transport of the wave energy to the medium on top of the surface. Additionally, precise control over the wavefield (*e.g.* wavelength, wave pattern, frequency) is possible through interdigital transducer (IDT) design and substrate selection [12]. These advantages have enabled studies that demonstrated unique interactions between SAWs and 2D materials, revealing quantum effects[13] such as acoustic exciton modulation [14], exciton transport [15] and phonon manipulation.[16] Despite their potential, the use of SAWs for actuating suspended 2D materials remains relatively unexplored. To our knowledge, with to date, only one study demonstrated ultrasound detection of shear horizontal surface waves across a resist layer.[17]

Once actuated, an effective readout method is needed to characterize the dynamic behaviour of suspended 2D materials. Electrical and optical readout methods are commonly used due to their ability to perform temporal measurements at high frequencies.[18] Electrical readout can detect changes in electrical properties, such as resistance or capacitance, occurring due to the membrane's dynamic response.[6,19,20] Optical readout, often employing laser-based techniques, detects the change in reflectivity while allowing for precise temporal measurements.[7,8] However, both methods have limitations in terms of spatial resolution since the readout signal is usually the weighted average of the variable (*i.e.* deflection or velocity) over a certain detection area. Electrical readouts offers resolution confined to the area of the electrodes, while optical readouts are limited to the optical focal spot size,[5] usually at the scale of a few μm. Although the resolution of these mapping techniques can be improved, the overall spatial detail cannot reach the sub-micron regime.[21] On the other hand, Scanning Probe Microscopy (SPM)-based readouts provide excellent spatial resolution, capable of capturing detailed topographical and dynamic features of the membrane at the nanoscale. In particular, atomic force microscopy (AFM) based methods excel in mapping localized displacement fields [22–24] and nanomechanical properties [25] of the materials with high spatial precision. As a drawback, SPM lacks the ability to capture real-time temporal dynamics at high frequencies due to the inertia of the scanning probe cantilever, and thus provides a time-averaged measurement. However, due to the non-linear tip-sample interaction, this method allows detection of standing wavefields with fixed nodal positions. In our study, we employ Atomic Force Acoustic Microscopy (AFAM) to map the SAW-induced standing wavefields on graphene, revealing acoustic field pattern variations that could not be discerned with conventional characterization techniques.

Here, we introduce SAW devices as a powerful tool for high-frequency on-chip mechanical actuation of suspended 2D material membranes. Specifically, we report the SAW excitation of suspended graphene layers at 375 MHz. We map the SAW-induced vibrating fields on graphene using AFAM and demonstrate reduced-wavelength localized acoustic fields across the graphene with respect to the piezoelectric substrate in the SAW delay line. Furthermore, we analyse the acoustic field maps and estimate the dispersion characteristics with effective bending rigidity and tension parameters using a tensioned plate model and experimentally obtained data. Our experimental observations are in very good agreement with COMSOL simulations in terms of acoustic pattern geometry and dimensions. This work is crucial for demonstrating the potential of SAW devices as a high-frequency on-chip actuation tool for suspended 2D materials, and it paves the way for localized field generation for nanoparticle manipulation applications, where smaller wavelength fields are required for more precise transport and patterning.

RESULTS AND DISCUSSION

*<u>AFAM mapping of vibrating SAW field</u>*

A 2-port SAW device was designed to produce surface waves of 10 μm wavelength on a lithium niobate (LiNbO3) substrate. Hence, inter-digitated transducers (IDTs) with a pitch of 5 μm were patterned and deposited on the surface by e-beam lithography and metal deposition steps (see Figure 1a). The detailed fabrication steps are summarized in the Methods section. To generate suspended 2D material membrane in the delay line, microcavities of 10 μm diameter and ~850 nm depth were formed on the center of the SAW device by focused ion beam (FIB) milling. Multilayer graphene flakes with thicknesses 9.4 nm to 37 nm were mechanically exfoliated and transferred on top of the cavities via viscoelastic stamps. An optical image of a graphene flake suspended over a FIB-milled cavity in a SAW device delay line is shown in the middle inset in Figure 1a.

The setup used for AFAM (Figure 1b) consists of a commercial AFM setup equipped with a lock-in amplifier and an external signal generator (SG). It enables concurrent mapping of the topography and acoustic wavefield from the signal collected at the photodiode (PD). During the AFAM measurement, a radio frequency (RF) signal at 375 MHz, amplitude-modulated at 10 kHz, is sent to one IDT port and generates a traveling wave that is reflected from the second port to create a standing wavefield in the transmission line between the IDTs. The right-hand side maps in Figure 1b show the imaged SAW field with the regular fringe pattern characterized by a half-wavelength (5 μm) pitch. The SAW field maps measured at different positions within the delay line exhibit the same features, confirming the homogeneity of the SAW field, and no significant difference is observed in the SAW field except at areas closer to the boundaries and outside the transmission line (see Supporting information S1.1 for maps acquired across various areas on the device).

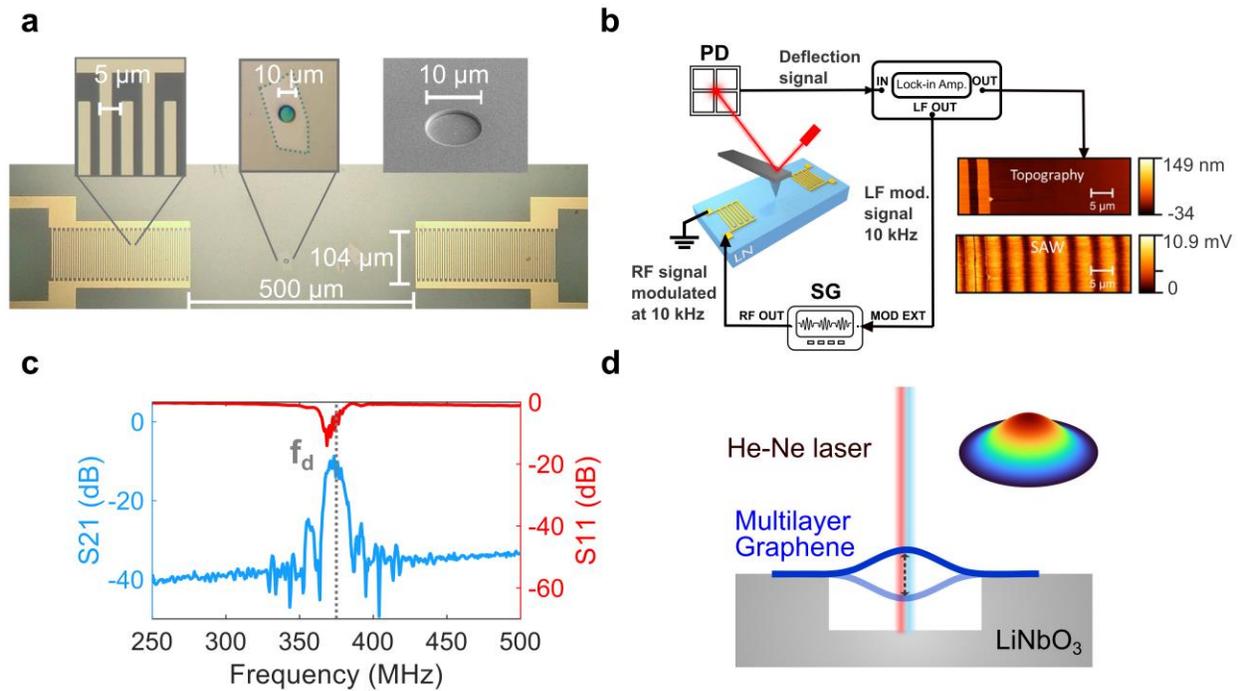

**Figure 1.** Experimental method: (a) Optical image of the SAW device with a suspended graphene membrane in the transmission line, with the SEM image of the microcavity. (b) Schematic of AFAM setup and mapping of a delay line region on bare LN device. (c) S-parameters of the SAW device with the drive frequency $f_d$, (d) Laser interferometry to detect fundamental frequency of suspended multilayer graphene on lithium niobate sample.

The SAW driving frequency fd is chosen based on the S-parameter measurement as 375 MHz, since the IDT generates SAW waves most efficiently at that frequency (Figure 1c). The fundamental resonance frequencies of the suspended graphene devices were determined via a He-Ne laser interferometer setup[26,27] where a 405 nm (pump) laser beam excites the graphene drum photothermally and the reflection of a 632 nm (probe) laser due to membrane vibration is captured by a photodetector (Supplementary data S2).

*Mechanical characterization*

For lower frequency dynamic characterization, fundamental resonances of the graphene drums were determined by laser interferometry in air (Figure 1d). Through simultaneous excitation and detection at varying frequencies (0-100 MHz), we extracted the dynamic behaviour of devices at a frequency band up to 100 MHz (see Supporting information S2). Specifically, amplitude and phase measurements over the frequency range, enabled us to detect the first two resonance frequencies $f_{01}$, $f_{11}$ of all devices. Third resonance frequency $f_{21}$, associated with mode (2,1) could also be detected for 4 out of 5 devices. Device data along with detected resonance peaks are given in Table 1.

**Table 1.** Device data and measured resonance frequencies

| Device | h [nm] | $\mu_\lambda$ [µm] | $f_{01}$ [MHz] | $f_{11}$ [MHz] | $f_{21}$ [MHz] |
|---|---|---|---|---|---|
| **D1** | 9.4 | 2.16 | 15.55 | 28.92 | 44.96 |
| **D2** | 14 | 2.28 | 18.23 | 32.38 | 51.99 |
| **D3** | 15 | 1.91 | 14.96 | 26.05 | 46.54 |
| **D4** | 34 | 1.83 | 17.53 | 28.62 | 47.53 |
| **D5** | 37 | 1.84 | 15.16 | 27.43 | - |

*SAW based actuation of suspended graphene membranes*

During SAW actuation, the surface wave propagates from the substrate surface first to supported and then to the suspended parts of the multilayer graphene flake, causing the graphene to vibrate out-of-plane. We imaged the resulting standing wavefield by de-modulation of the AFM deflection signal (see Methods for details). To determine the spatial variation in acoustic amplitudes, we performed measurements on the bare LiNbO$_3$ substrate and on both supported and suspended parts of graphene flakes of different thicknesses.

To ensure a homogeneous field entering the suspended graphene membranes, we positioned the cavities in the LiNbO$_3$ near the centerline of the area between the two IDTs. We subsequently assessed whether the cavity played a role in modulating the SAW field by acquiring AFAM maps across an uncovered microcavity (Figure S1.2). Owing to the smooth surface and edges of the cavity, no significant changes in SAW field due to the cavity were observed. This demonstrates that acoustic wave fields are not significantly altered by the cavity.

To analyze how the acoustic wavefield is affected by graphene layers of different thicknesses, we fabricated 5 suspended graphene drums with varying thicknesses (9.4 nm, 14 nm, 15 nm, 34 nm, 37 nm) on the delay line of identical SAW devices. The vibration field on the suspended part of the graphene was mapped under the same actuation conditions (375 MHz, 15 dBm) for each device. The topography and AFAM wavefield for three devices with thicknesses of 9.4 nm, 15 nm and 34 nm is given in Figure 2 and the results from the remaining two devices are given in the Supporting information S1.3. Height profiles taken across the graphene drums showed downwards bending of the suspended flake towards the cavity floor, with the level of bending decreasing as the thickness increased (Figure 2a). Upwards bending was also observed for thicker flakes (~34 nm). The SAW field maps (Figure 2b) clearly showed wavefields with smaller wavelengths and different circular patterns on suspended graphene compared to the supported parts. The 10 μm wavelength on supported graphene reduces to a wavelength of 1.83-2.28 μm on the suspended graphene, corresponding to a reduction by more than a factor 4. We note that this 76% reduction in the wavelength could not be captured in non-contact optical measurements with μm-level spatial resolutions such as laser doppler vibrometry (See Supporting information S6). Multiple profiles across the wavefield were extracted from the mapped vibration data for each device, and the distance between consecutive peaks, corresponding $\lambda_{exp}/2$, was measured (Figure 2c). Peak distances were converted into wavelength and wavenumber. Mean and standard deviation of the wavelength ($\mu_\lambda, \sigma_\lambda$) and wavenumber ($\mu_k, \sigma_k$) were calculated directly from the corresponding dataset (S3).

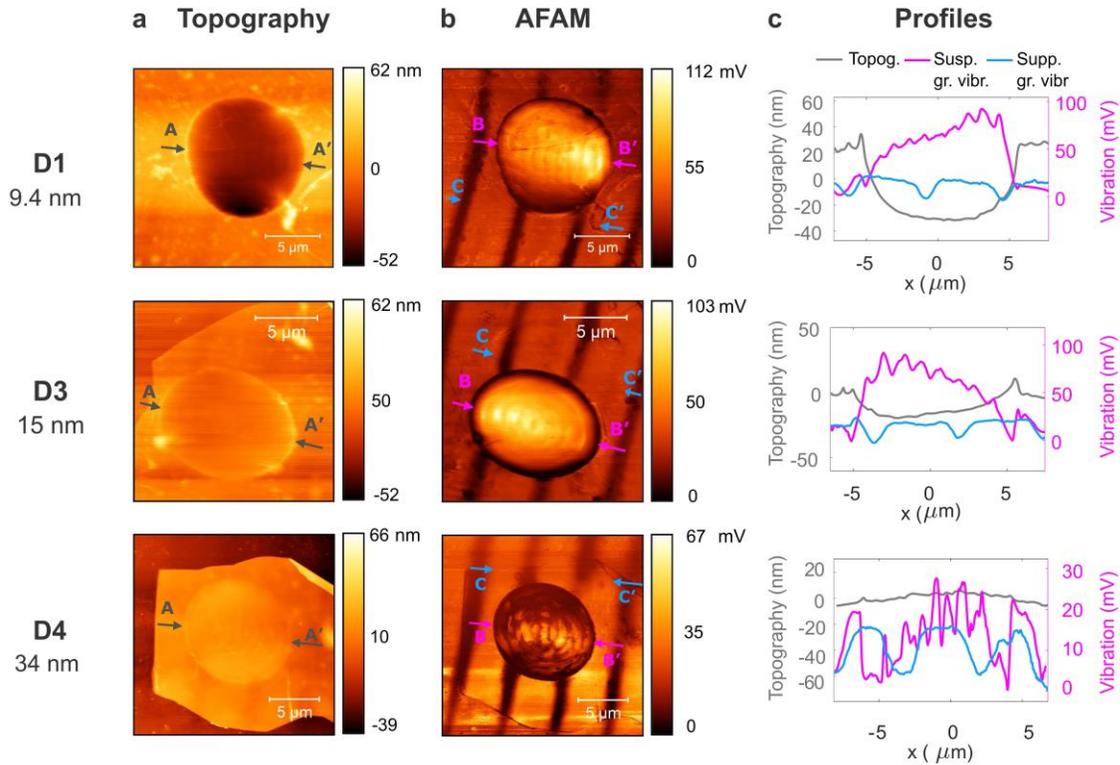

**Figure 2.** AFAM maps across suspended graphene membranes. (a) Topography image and (b) AFAM image of the vibrating field on suspended graphene drums and supported graphene on LN substrate. (c) Acoustic wave and topography line profiles across the propagation direction. Profiles are obtained from the lines passing through the points specified by grey (AA'), pink (BB') and green (CC') locations in (a) and (b).

*Pre-tensioned plate model*

When a surface wave impinges on suspended graphene, the Rayleigh type wave transforms into a flexural wave on the thin layer. Rayleigh waves have elliptic particle motion composed of longitudinal and transverse components, and propagate through the surface of the solid. In contrast, flexural waves involve bending deformation primarily in thin structures such as plates or beams, with motion that includes transverse displacements and rotations, propagating along the structure's length. In order to model this effect, the multilayer graphene flakes used in this work were modelled as pre-tensioned plates, taking both bending rigidity (D) and tension (T) into consideration. The dispersion relation for flexural waves in the pre-tensioned plate model is given as (1), where $\omega_k$ is the angular frequency at wavenumber $k$ and $\rho h$ is the areal mass density of graphene.

$$\omega_k = \sqrt{\frac{D}{\rho h}k^4 + \frac{T}{\rho h}k^2} \qquad (1)$$

Generally, for thin flakes (<20 layers), tension dominates over bending rigidity, leading to membrane-like behavior where the wave speed is determined by tension. As indicated by equation (1) for pre-tensioned plates, in lower frequency modes, the effect of bending rigidity decreases with lower wavenumber, hence the effect of tension can be observed. However, due to the cubic dependence of bending rigidity on thickness, thicker flakes will deviate from this behavior faster with increasing frequency and wavenumber.

As the first (2-3) resonance frequencies of the devices are detected by laser interferometry, and AFAM detects the response to higher frequency SAW excitation, it is possible to estimate the effective bending rigidity parameter $(D/\rho h)_{eff}$ and effective tension parameter $(T/\rho h)_{eff}$ from experimentally obtained data. Wavenumbers corresponding to the first 3 resonance peaks can be found as $k_{01} = 3.1962/R$, $k_{11} = 4.6109/R$, $k_{21} = 5.9057/R$ for clamped circular plates with radius R. For the high frequency data, wavenumber was obtained from measured wavelength using equation (2):

$$k_{exp} = 2\pi/\lambda_{exp} \qquad (2)$$

By knowing the frequency and wavenumber, the phase velocity at each frequency can be found using the relation $c_p = \omega/k$ at different frequency points. From the experimental data, it is observed that the phase velocity increases from ~160 m/s at the fundamental frequency to ~700 m/s at the SAW drive frequency of 375 MHz. Compared to the SAW phase velocity in the lithium niobate (~3750 m/s), the suspended graphene leads to a local decrease in wave speed governed by the change of effective stiffness of the medium.

A fitting procedure is performed with the laser interferometer and AFAM data based on equation (1). Figure 3a shows the detected peak positions as the frequencies of the modes (0,1), (1,1) and (2,1), which are represented in Figure 3b as diamond, square and triangle markers respectively. For device D5, which does not have an apparent $f_{21}$ peak at its spectrum, the fit was performed with 3 data points instead of 4. Figure 3b shows the fitted dispersion curves for three of the devices with 9.4 nm, 14 nm, 34 nm graphene thicknesses. The circles in the panel b represent the mean value of the wavenumber ($\mu_k$) at SAW drive frequency 375 MHz, and the standard deviation ($\sigma_k$) is given as horizontal error bars. Figure 3c shows the insets from Figure 3b where the lower frequency modes are marked. Each color represents a device with a given graphene thickness. Experimental data points fit well with the estimated dispersion curves, with the goodness parameter R2>0.99 for all devices. Data fitting results for each device are given in the Supporting Information (S4).

Figure 3b clearly illustrates the quadratic dependence of frequency on the wavenumber. A quadratic dispersion is indicative of plate-like behavior dominated by bending rigidity. This observation aligns with the tensioned plate model (eqn. 1), in which the tension term scales with $k^2$ while the bending rigidity term scales with $k^4$ in the frequency-squared $\omega^2$ relation. Consequently, as the frequency and wavenumber increase, the contribution of the tension term diminishes relative to the dominant bending rigidity effect. This trend underscores the increasing significance of bending rigidity in governing the system's dynamic response at higher frequencies and wavenumbers, especially for thicker layers.

From the fitting procedure, we obtained estimates for the effective bending rigidity $\left(\frac{D}{\rho h}\right)_{eff}$ and tension parameters $\left(\frac{T}{\rho h}\right)_{eff}$, which correspond to the bending rigidity and tension divided by the areal mass density of the multilayer graphene, respectively. The fitted effective bending rigidity values are notably higher for devices D1 and D2 compared to both the remaining devices and the analytical estimation (Supplementary Information S4). In contrast, the fitted results for devices D3, D4, and D5 are closely aligned with each other and are more consistent with the analytical estimates obtained from $\frac{D}{\rho h} = \frac{Eh^2}{12\rho(1-v^2)}$ assuming Young's modulus E=1 TPa and Poisson's ratio $v$=0.165. The observed discrepancies between the analytical estimates and the experimental fits highlight the complex and intriguing nature of 2D material dynamics, where thinner layers are more susceptible to multiple external influences. Variations in apparent stiffness may arise from fabrication and flake-related defects as well as from tip-induced forces during the AFM scan, underscoring the sensitivity of these measurements to external and device-specific factors.

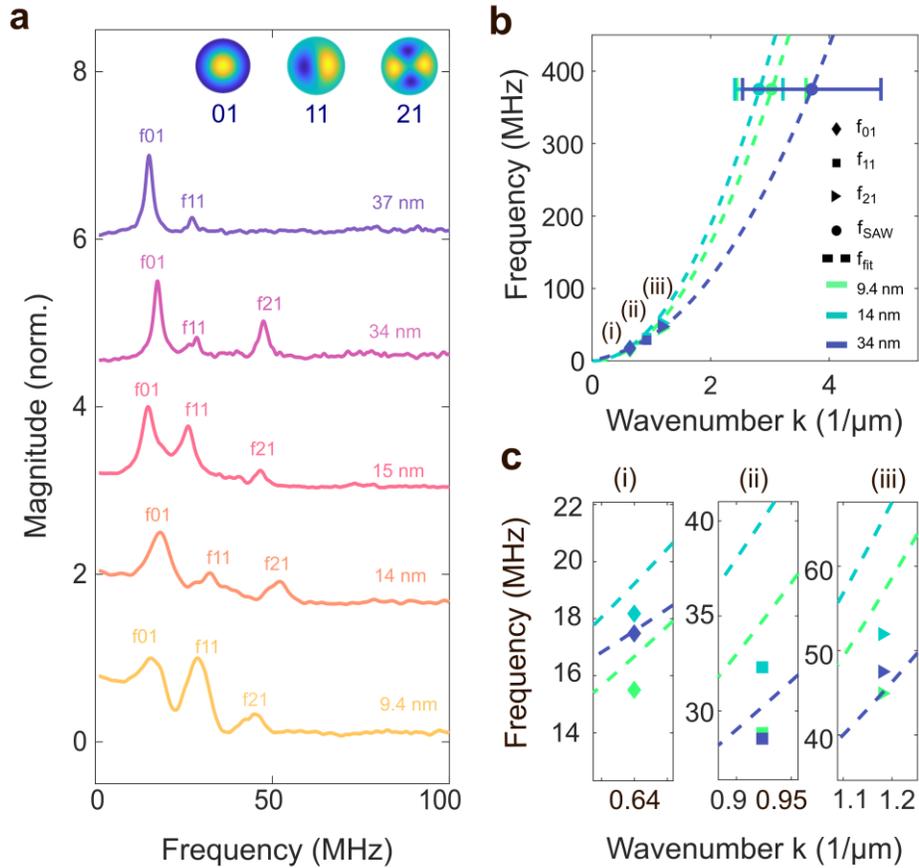

**Figure 3.** Wave dispersion curves for the pre-tensioned plate model of devices D1,D3,D4 with experimentally obtained values. (a) Lowest resonance frequencies detected for each device from laser interferometry measurements. (b) Fitted dispersion curves for D1, D2 and D4. (c) Insets (i),(ii),(iii) from (b) around first three resonance frequencies.

*Simulation of SAW-induced wavefields on suspended graphene membranes*

COMSOL Multiphysics software was used to create a 3D model of the suspended graphene drum structure (See Supporting Information S3). The Young's modulus (E) and pretension (T) values obtained from the fitted bending rigidity were inserted into the model, and the vibration response was simulated. Figure 4a and 4b show the simulated and experimentally mapped wavefields on suspended graphene at 375 MHz for D4 (h=34 nm) respectively. A comparison of the numerical and experimental data over the profiles taken between the white, dashed arrows in panels (a) and (b) can be seen in Figure 4c. In this figure, absolute out-of-plane displacements were normalized between 0 to 1. From the figure, we clearly observe that numerical estimation and experimental measurements align well, and thus the model can be used to predict wavefields on suspended graphene layers, with more closely matching results expected for relatively thicker and flatter topography). Our model also applies to different surface geometries (*i.e.*, non-circular cavities), as well as different actuation conditions (*i.e.*, wavelength, frequency). To demonstrate this, we fabricated both non-circular (square) and circular cavities with varying diameters to perform further measurements. The results of these measurements, along with the corresponding simulation results, are provided in Supporting Information S6. All cavities were covered by the same large flake, allowing for consistent comparison across different cavity design parameters. We observed the effects of changing the cavity size and shape on the system's response. By varying the cavity diameter, we were able to alter the number of nodal lines in the suspended part. Furthermore, by altering the geometry to a square cavity, we generated linear nodal lines perpendicular to the SAW wave pattern, highlighting the impact of cavity geometry. Simulated and experimentally measured wavefields for the square (Fig 4 d,g), and circular cavity with 5 μm diameter (Fig 4 e,h) and 4 μm diameter (Fig 4 f,i) are given below. This demonstration shows the applicability of the method for varying cavity sizes and geometries to design and control the wavefield modulation for further applications. Importantly, simulations and experiments for additional geometries match well also when using general values for the material properties of graphene (E=1 TPa), highlighting the predictive power of the model. Furthermore, this approach allows not only for tuning the wavelength of the waves on the suspended membrane but also for bending or refracting the waves at the interface, analogous to how a lens manipulates optical waves.

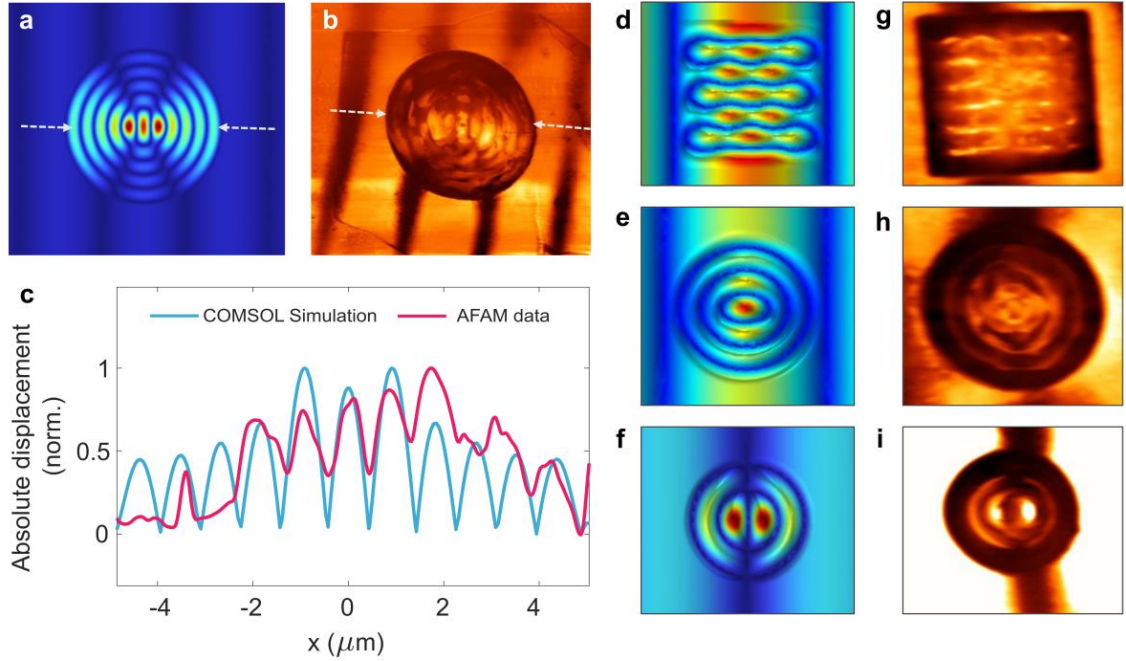

**Figure 4.** Comparison of the simulation and the experimental data for device 4 (h=34 nm suspended graphene). (a) COMSOL simulation for a suspended circular graphene membrane with h=34 nm, with $\left(\frac{D}{\rho h}\right)_{eff}$ and $\left(\frac{T}{\rho h}\right)_{eff}$ obtained from data fitting. (b) AFAM mapping of the wavefield on device 4. (c) Comparison of normalized simulation and experimental data over the profiles defined between the white dotted arrows in (a) and (b). (d-f) Simulation results for a 23 nm thick graphene flake suspended over square and smaller circular cavities, (g-i) AFAM mapping of the vibration field on these devices.

CONCLUSION

Here we present a SAW-based platform for on-chip mechanical actuation of suspended 2D materials and mapping of acoustic fields on suspended graphene layers at very high frequencies. By using atomic force acoustic microscopy, we are able to detect and spatially map high frequency (375 MHz) flexural vibrations on suspended graphene layers with different thicknesses. Combining laser interferometry and AFAM, we demonstrate the dispersive behaviour of the flexural wave on suspended graphene layers and characterize their dynamic response with a tensioned plate model.

With the proposed actuation platform, it is possible to spatially modulate the SAW wavefields by means of suspended 2D materials. We demonstrate that SAW wavefields, which typically have fixed wavelengths, can be locally transformed into smaller wavefields confined within the suspended membrane area. With the current device structure, we observed a reduction of a 10 µm SAW wavelength to shorter wavelengths between 1.83 µm and 2.28 µm, corresponding to a reduction greater than 76%. Such modulations can find future applications in nanomaterial transport studies where ability to design and deterministically locate such smaller-wavelength regions in a homogenous field would be a crucial step towards higher precision on-chip manipulation. As the wavelength is reduced, mapping such wavefields becomes a challenge since non-contact optical measurements with µm-level spatial resolutions (*e.g.* laser doppler vibrometry) cannot capture such small-scale vibrations (See Supporting information S6). Hence, the requirement for high spatial resolution emphasizes the need for probe-based techniques such as AFAM. In summary, the combined AFAM-SAW platform is poised to reveal novel functionalities and phenomena enabled by acoustic-matter interactions at the thin membrane limit. It can be used to guide, refract and reflect acoustic waves on a surface, similar to the way lenses guide optical waves. In particular, the dynamic strain field and electric field can propel new applications requiring the close interplay of phonons, photons and electrons in next-generation acousto-opto-electronic nanodevices.

METHODS

*Design:* The 2-port SAW device was designed with a $\lambda/4$ fingerwidth to generate a $\lambda=10$ μm wavelength surface waves. The number of fingerpairs for each IDT port was determined to be 31 to match the electrical impedance of the device to the 50 $\Omega$ output impedance of the source equipment in order to minimize transmission losses.

*SAW device fabrication:* 128° Y-cut X-propagating Lithium Niobate (LN) wafers of 500 μm thickness were purchased from Siegert Wafer GmbH (Germany) and diced into 10 mm x 5 mm chips using a Disco DAD dicing saw (Disco Hi-Tec Europe GmbH, Germany). LN chips were first spin coated with PMMA A3 (495) and then with PMMA A6 (950) resists at 4000 rpm. After each spin coating, chips were baked at 115°C on a hotplate for 2 minutes. As a final layer, AR-PC 5090 (Electra 92) was spin coated at 4000 rpm, and the chips were baked at 90 °C for 1 minute. IDTs and electrodes were patterned on the chip via electron (e-) beam lithography (EBPG-5000+, Raith GmbH, Germany). After e-beam lithography, the chips were developed in water (60 s), pentyl acetate (30 s), MIBK:IPA (30 s), and IPA (30 s), respectively. In the following step, Ti (5 nm) and Au (80 nm) layers were deposited using electron beam evaporation (Temescal FC-2000, Ferrotec Europe GmbH, Germany), and the process was finalized with a lift-off in 70 °C anisole to remove excess resist. The fabricated SAW devices were connected to the PCB by wirebonding.

*Microcavity formation:* We employed Focused Ion Beam (FIB) milling to fabricate microcavities of 10 μm diameter. The resulting surface quality is important for the fabrication process as roughness induces additional effects such as wrinkles on suspended flakes and poor sticking during transfer. Additionally, surface roughness at the cavity edges can amplify reflections and change the SAW field impinging on the suspended membrane. To prevent the roughness on the cavity edges and unwanted effects of charge accumulation due to the ion beam, a 20 nm chromium layer was first deposited on the SAW device surface to distribute the charge on surface. FIB milling (FIB/SEM Helios G4 CX, Thermo Fisher Scientific) with a Ga+ ion source of 30 kV acceleration voltage and 1.1 nA beam current was used to make 10 µm-diameter microcavities around the center of the SAW delay line. After FIB milling the remaining copper layer was etched using a chromium etchant (composed of perchloric acid ($HClO_4$), and ceric ammonium nitrate ($Ce(NH4)_2(NO_3)_6$) for 20 s). This way, copper was removed without any significant damage to the electrodes. Characterization of the cavity and surface topography was done with standard AFM contact mode scanning.

*Graphene transfer:* Mechanically exfoliated graphene flakes were transferred to the SAW substrate delay line using the viscoelastic stamping method [28]. First, the graphene flakes were exfoliated on a $SiO_2$/Si wafer. Then, exfoliated flakes were imaged under an optical microscope to assess the size and thickness. Selected flakes were picked up by a viscoelastic stamp made of PDMS bubble and placed on top of the cavity on SAW delay line. After the transfer, the flake topography was determined by AFM to characterize the initial static condition of the structure. This is important to know beforehand since the transfer process may result in wrinkles on the flake and bending (upwards or downwards) which may impact acoustic wave propagation. To characterize this initial state, AFM contact mode imaging was used.

*SAW device characterization:* The dynamic behavior of the SAW device was characterized by the scattering parameters obtained using a 2-port Vector Network Analyzer (VNA). Each port was connected to an IDT via SMA connectors on the PCB. A band-limited AC signal (250–500 MHz) was sent to one of the IDTs, and the signal was received by the other IDT. Through the analysis of the reflection (S11) and transmission (S21) parameters over the given frequency range, the resonance frequency (*i.e.*, minimum reflection and maximum transmission) was determined and subsequently used as the driving frequency for AFAM measurements. A relatively wide resonance range around 375 MHz was observed and this frequency was selected as driving frequency for the SAW device.

*Probing the SAW field:* AFAM scans were acquired in contact mode with a Bruker AFM equipped with a lock-in amplifier. One port of the SAW device was given an RF signal at the device's driving frequency, modulated at 10 kHz at 15 dBm input power. This was done for each device scan. The modulation signal was supplied by the lock-in amplifier and used again as a reference signal for the demodulation of the deflection signal detected by the AFM photodiode. Through de-modulation, the resulting SAW-displacement field was extracted from the topography data. The Gwyddion software was used to analyze the obtained data.

**AUTHOR INFORMATION**

**Corresponding Authors**

*E-mail: s.caneva@tudelft.nl

g.j.verbiest@tudelft.nl

**Author Contributions**

The manuscript was written through contributions of all authors. All authors have given approval to the final version of the manuscript.

**Notes**

The authors declare no competing financial interests.

## ACKNOWLEDGMENTS

H. N. A, and S.C. acknowledge funding from the European Union's Horizon 2020 research and innovation program (ERC StG, SIMPHONICS, Project No. 101041486). X.Y. acknowledges funding from the Chinese Scholarship Council (Scholarship No. 202108270002). We acknowledge A. Keşkekler and F. Alijani for the use of laser interferometry setup.

**Supporting Information Available** Additional data and results mentioned in the manuscript

Table of content figure

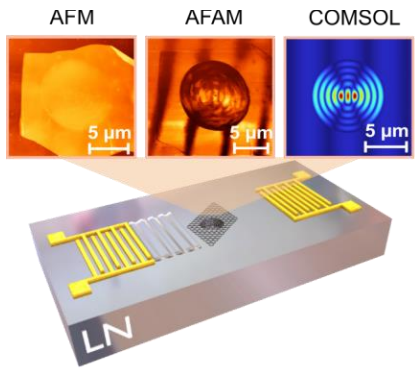